\documentclass[doublecol,figures]{epl2}
\usepackage{graphicx}
\usepackage{amsmath,amssymb}

\title{Intrinsic Ratchets}

\author{M. van den Broek\inst{1}\thanks{E-mail: \email{vandenbroek.martijn@gmail.com}} \and R. Eichhorn\inst{2}  \and C. Van den Broeck\inst{1}}
\shortauthor{M. van den Broek \etal}

\institute{                    
\inst{1} Hasselt University - B-3590 Diepenbeek, Belgium\\
\inst{2} Fakult\"at f\"ur Physik, Universit\"at Bielefeld - 33615 Bielefeld, Germany
}
\pacs{05.40.-a}{Fluctuation phenomena, random processes, noise, and Brownian motion}
\pacs{05.70.Ln}{Nonequilibrium and irreversible thermodynamics}
\pacs{62.25.Fg}{High-frequency properties, responses to resonant or transient (time-dependent) fields}

\abstract{
We present a generic formalism to describe Brownian motion of particles with intrinsic asymmetry and give predictions for the drift behavior in unbiased time-dependent force fields. 
Our findings are supported by molecular dynamics simulations.}

\begin{document}

\maketitle

\section{Introduction}

In view of their potential applications in bio- and nanotechnology, Brownian motors have in recent years been the object of intensive research \cite{astumian, julicher,reimann,marchesoni}. As is well known, the rectification of the thermal motion of Brownian particles involves the breaking of underlying symmetries. On the one hand, the system has to operate under nonequilibrium conditions to break the microscopic equilibrium symmetry of detailed balance.  Spatial symmetry on the other hand is usually broken by applying asymmetric external forcing. The two most cited paradigms in this context are the flashing and rocking ratchets (see, \textit{e.g.}, \cite{hanggi}), in which an external space-periodic but space-asymmetric forcing using a ratchet-like potential is applied. Somewhat surprisingly, the case in which an inherent asymmetry of the Brownian particle itself provides the spatial asymmetry has not been discussed in the context of periodic forcing. We will refer to such Brownian motors as intrinsic ratchets. In this letter, we will introduce and solve the equations of motion that generically describe this type of thermal rectification. 

\section{Generic equations of motion}

The motion of a Brownian particle (speed $v$ and mass $M$) is usually described by the following Langevin-Newton equation:
\begin{equation}
M \frac{\upd v}{\upd t}= -\gamma v +F+ \xi.
\end{equation}
Here $\gamma$ is the friction coefficient, $F$ is an applied external force and $\xi$ a Gaussian white noise, whose intensity is determined by the fluctuation-dissipation relation. Equivalently, one can write the Fokker-Planck equation for the probability distribution $P(v)$ of the speed, namely,
\begin{equation}
\partial_t P= \partial_v \left(\frac{\gamma}{M} -\frac{F}{M}+\frac{\gamma k_B T}{M^2}\partial_v \right)  P,
\end{equation}
where $k_B$ is Boltzmann's constant and $T$ is the temperature of the
bath particles.  The above equations can be derived from a microscopic
description by assuming that the mass of the Brownian particle is much
larger than that of surrounding particles. They form the starting
point for deriving the properties of flashing or rocking ratchets. In
fact, since one needs to apply spatially asymmetric forcing in these
systems, the analysis is typically carried out at the simpler level of
overdamped motion. The latter provides a closed description in terms
of the position variable only and is known to be a very good
approximation in most situations. As we will see below, we however do
not need spatially dependent forcing for the rectification in
intrinsic ratchets. This greatly simplifies the analysis, even at the
underdamped level. Indeed, when the forcing $F$ is position
independent, the stochastic variable $v$ is Gaussian, and it suffices
to study the equations of motion for the first two moments of the
velocity. By choosing as units of time, velocity and force the
relaxation time $\tau_r = M/\gamma$, the thermal speed $v_T =
\sqrt{k_B T/M}$ and $\gamma v_T$, the following equations are obtained
for the moments $v_1=\langle v \rangle$ and $v_2=\langle v^{2}
\rangle-1$:
\begin{align}
\frac{\upd v_1}{\upd t} 
&= - v_1 + f,
\nonumber\\
\frac{\upd v_2}{\upd t} &= - 2 v_2 + 2 f  v_1.
\end{align}

We now argue that a minor modification of these equations describes the case of intrinsic ratchets. We first note that the possible asymmetry of the Brownian particle does not appear in the above equations, basically because the relaxation is described by linear response. As a result, the equation for the first moment, which is the central object of interest, is not coupled to the second moment.  The asymmetry of the particle will appear at a next order of perturbation, at the level of nonlinear relaxation. Furthermore, the resulting correction appearing in the equation for $v_1$ still has to vanish when operating under equilibrium conditions, \textit{i.e.}, when  $v_2=0$. The simplest analytical correction is thus a term of the form $\alpha v_2$, where the constant $\alpha$ quantifies the strength of the asymmetry. Since this term acts like a perturbation on the first moment we can dismiss, to lowest order, the correction that will appear in the equation for $v_2$. The intrinsic ratchet is thus described at lowest order (with $\alpha$ effectively playing the role of a small dimensionless parameter) by the following generic set of equations:
\begin{align}
\frac{\upd v_1}{\upd t} &= - v_1 + \alpha v_2 + f, \nonumber \\
\frac{\upd v_2}{\upd t} &= - 2 v_2 + 2 f  v_1.
\label{equationmotion}
\end{align}

In addition to the above handwaving arguments, we note that the
equations of motion given in eq. (\ref{equationmotion}) can be derived
from microscopic theory of a Brownian particle moving in a bath of an ideal gas,
by an expansion in the ratio of the mass of
the gas particles ($m$) over the mass of the Brownian particle ($M$)
\cite{vandenbroeckmotor, meurs, martijn1}. Such a
derivation also provides explicit expressions for the open parameters
$\alpha$ and $\gamma$ (or $\tau_r$) behind eq.~(\ref{equationmotion})
in terms of microscopic quantities.
Concrete examples for the cases of translational and rotational motion
of an asymmetric object suspended in a thermalized gas will be given below.

In the remainder of this letter, we focus on the rectification,
\textit{i.e.}, the appearance of a non-zero average drift velocity,
when the particle is subjected to an unbiased time-periodic force
$f(t)$. This scenario is the analogue of the rocking ratchet for
particles with intrinsic asymmetry.

\section{Piecewise constant forcing}

Eq.~(\ref{equationmotion}) with time-periodic forcing $f(t)$ has a mathematical structure similar to the Newton equation of motion for a parametric oscillator. It is therefore out of the question to find a general analytical solution. Instead we  turn to the case of piecewise constant forcing (square wave profile), \textit{viz.},
\begin{align}
0 \le t < \tau/2 : \quad &f(t) = f_0, \nonumber \\
t \le \tau/2 < \tau : \quad &f(t) = -f_0.
\label{force}
\end{align}
Introducing the vector notation
\begin{equation}
V(t) = \begin{bmatrix}
v_1\\
v_2
\end{bmatrix} (t),
\end{equation}
one readily finds the solutions in the separate time regimes, with $f = f_0$ and $f = -f_0$ respectively:
\begin{align}
V_+(t)
&= A_+(t) C_+ + B_+,
\label{sol1}
\\
V_-(t)
&= A_-(t)
C_- + B_-.
\label{sol2}
\end{align}
The time-propagators $A_{\pm}$ are given by:
\begin{equation}
A_{\pm}(t) = 
\begin{bmatrix}
\pm \frac{1-d_{\pm}}{4 f_0} e^{-(3 + d_{\pm}) t / 2}   & 
\pm \frac{1+d_{\pm}}{4 f_0} e^{-(3 - d_{\pm}) t / 2} \\
e^{-(3 + d_{\pm}) t / 2} & 
e^{-(3 - d_{\pm}) t / 2}
\end{bmatrix},
\end{equation}
with $d_{\pm} = \sqrt{1 \pm 8 \alpha f_0}$.
$B_{\pm}$ are the steady state solutions:
\begin{equation}
B_{\pm} = 
\begin{bmatrix}
\pm f_0 / (1 \mp \alpha f_0)\\
f_0^2 / (1 \mp \alpha f_0)
\label{steadystate}
\end{bmatrix}.
\end{equation}

\begin{figure}
\onefigure[width=0.6\columnwidth]{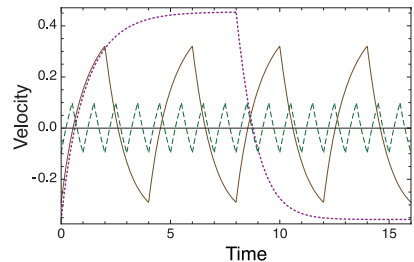}
\caption{
Time evolution of the first moment of the velocity $v_1$ (in thermal speed $v_T$ units)
subject to a modulating square force with periods 1, 4, and 16 (dashed, solid, dotted curve).
The relaxation time $\tau_r$ of the particle is the time unit.
The particle's asymmetry is $\alpha = 0.3$ and force amplitude is $f_0 = 0.4$ (in units $M v_T / \tau_r$, with $M$ the particle's mass).
}
\label{fig:velocity}
\end{figure}

The vector constants $C_\pm$ are specified by the assumption that we operate in the steady state regime, hence the velocity moments in eqs. (\ref{sol1}) and (\ref{sol2})  satisfy time-periodic boundary conditions, $V_+(0) = V_-(\tau)$. This, together with continuity at $t = \tau/2$, $V_+(\tau/2) = V_-(\tau/2)$, leads to a solution,
\begin{align}
C_+ &= (A_2^{-1} A_1 - A_4^{-1} A_3)^{-1} (A_2^{-1} - A_4^{-1}) (B_- - B_+),\\
C_- &= (A_1^{-1} A_2 - A_3^{-1} A_4)^{-1} (A_1^{-1} - A_3^{-1}) (B_+ - B_-),
\end{align}
with $A_1 = A_+(\tau/2), A_2 = A_-(\tau/2), A_3 = A_+(0), A_4 = A_-(\tau)$.
We will not reproduce here the resulting expression for the time-dependent average speed $v_1$. It is extremely cumbersome, and, strictly speaking, only valid to lowest order in the asymmetry contribution $\alpha$. For an illustration of the typical time dependence of $v_1$ we refer to fig.~\ref{fig:velocity}.   
We remark that our theory yields accurate results when $\alpha f_0 \ll 1$. Under the described ratchet operation, the speed $v_1$ and second moment $v_2$ are then of order $f_0$ and  $f_0^2$ respectively. This means that the nonlinear correction term $\alpha v_2$ in the equations of motion [eq.~(\ref{equationmotion})] is a factor $\alpha f_0$ smaller than the linear $v_1$ term.

\begin{figure}
\onefigure[width=0.7\columnwidth]{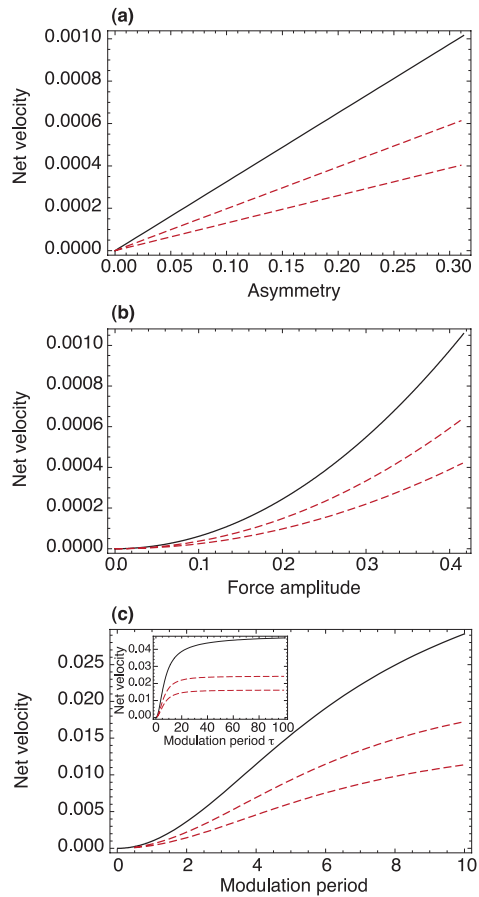}
\caption{Time-average net velocity $v_{\text{net}}$ as a function of
\textbf{(a)} the asymmetry $\alpha$ of the Brownian particle,
\textbf{(b)} the amplitude $f_0$ of the modulating force, and
\textbf{(c)} the modulation period $\tau$.  Exact analytical
solutions for square modulation -- solid curves -- coincide, for the
presented parameter range, with the first order approximation
[eq.~(\ref{lo})]. Dashed curves represent results for harmonic
modulation (upper curves) and sawtooth modulation (lower curves),
eqs. (\ref{loh}) and (\ref{los}), obtained from first order
perturbation theory. Numerical integration results can not be distinguished from
the analytical solutions in the shown graphs. If not in the abscissa, parameter values
are $\alpha = 0.3$, $f_0 = 0.4$, $\tau = 1$.  Units for velocity,
time and force are the particle's thermal speed $v_T$, relaxation
time $\tau_r$ and $M v_T / \tau_r$ (mass $M$) respectively; $\alpha$
is dimensionless.}
\label{fig:netvelocity}
\end{figure}

The quantity of central interest is the resulting time-average net speed, being the average net displacement over a period $\tau$ divided by this period: 
\begin{equation}\label{vnet}
v_{\text{net}} = \tau^{-1} \int_0^{\tau} v_1(t) \, \upd t.
\end{equation}
Again, the exact expression for $v_{\text{net}}$ is extremely long. In any case, our approach is limited to small $\alpha$, so it suffices to reproduce the lowest order term in  $\alpha$:
\begin{equation}\label{lo}
v_{\text{net}} \simeq \alpha f_0^2 \left(1 - \frac{4}{\tau} \tanh \frac{\tau}{4} \right).
\end{equation}
We note that the next term in the expansion in $\alpha$ is an order of magnitude $(\alpha f_0)^2$ smaller.

We make the following observations.
First, there is no directed motion, $v_{\text{net}} = 0$, in the absence of forcing, $f_0 = 0$, or when the particle has intrinsic symmetry, $\alpha = 0$. 
Second, $v_{\text{net}}$ is an uneven function of $\alpha$, hence an inversion of the asymmetry, $\alpha \rightarrow -\alpha$, results in an inversion of the speed of net motion. 
We represent $v_{\text{net}}$  as a function of the asymmetry, $\alpha$, the amplitude of the force, $f_0$, and the period, $\tau$, in fig.~\ref{fig:netvelocity} (solid curves). In all three cases, the lowest order approximation, eq.~(\ref{lo}), is in fact indistinguishable from the exact result for the chosen range of values of $\alpha$ and $f_0$. Finally we note that the maximum speed $v_{\text{net}}^{\text{lim}}\simeq \alpha f_0^2$  is reached in the limit of very slow modulation, $\tau \rightarrow \infty$. Since this speed is expressed in units of thermal velocity, we conclude that one can reach high net drift speeds, comparable to thermal speeds, by applying unbiased  periodic forcing of small to moderate intensity to intrinsic ratchets.
\section{Other periodic forcings}
\begin{figure}
\onefigure[width=0.6\columnwidth]{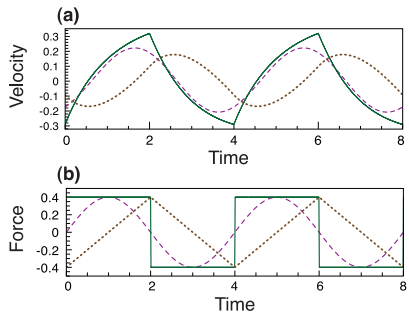}
\caption{
Time evolution of
\textbf{(a)} the first moment of the velocity $v_1$ (in thermal speed $v_T$ units)
for the periodic forcings in \textbf{(b)} (correspondence by line style).
For harmonic and symmetric sawtooth profiles the first order approximation
[eq.~(\ref{v1alpha})] is used, for square modulation the full
analytical solution is shown. In all three cases, a numerical solution of
the dynamic equation is indistinguishable from the analytical results in the graph.
Parameter values are period $\tau = 4$, force amplitude $f_0 = 0.4$ and asymmetry $\alpha = 0.3$.
Unit of time is the particle's relaxation time $\tau_r$, unit of force is $M v_T / \tau_r$, with $M$ the particle's mass.}
\label{fig:otherforcings}
\end{figure}

To investigate the effect of other types of periodic forcing, such as
harmonic or symmetric sawtooth [\textit{cf.}
fig.~\ref{fig:otherforcings}(b)],
we resort to a perturbational solution of
eq.~(\ref{equationmotion}).
As the contribution due to the intrinsic asymmetry $\alpha$ is considered small, we can make the
following first-order ansatz for the velocity moments:
\begin{align}
v_1 & = v_{1,0} + \alpha v_{1,1} \, ,
\label{v1alpha} \\
v_2 & = v_{2,0} + \alpha v_{2,1} \, .
\label{v2alpha}
\end{align}
With this ansatz, eq.~(\ref{equationmotion}) can be solved
to first order in $\alpha$ for arbitrary periodic force fields $f(t)$,
yielding the steady state solutions
\begin{align}
v_{1,0}(t) & = \int_{-\infty}^t \upd t' \, e^{-(t-t')} f(t') \, ,
\label{v10} \\
v_{2,0}(t) & = \int_{-\infty}^t \upd t' \, e^{-2(t-t')} 2f(t')v_{1,0}(t') \, , 
\label{v20} \\ 
v_{1,1}(t) & = \int_{-\infty}^t \upd t' \, e^{-(t-t')} v_{2,0}(t') \, ,
\label{v11} \\
v_{2,1}(t) & = \int_{-\infty}^t \upd t' \, e^{-2(t-t')} 2f(t')v_{1,1}(t') \, .
\label{v21} 
\end{align}
Using $f(t+\tau)=f(t)$ it is easy to show that these expressions are
indeed periodic with periodicity $\tau$. 
The results for the time evolution of the
first moment $v_1$ under harmonic or sawtooth forcing are reproduced in
fig.~\ref{fig:otherforcings}(a); they are indistinguishable from numerically integrated
solutions of the original eq.~(\ref{equationmotion}).
For comparison, the results for the
square wave profile are also included. 

With regard to the net velocity $v_{\text{net}}$ as defined in eq.~(\ref{vnet}),
we observe that for unbiased symmetric forcings
$f(t+\tau/2)=-f(t)$, and thus
$\int_0^\tau \upd t' \, v_{1,0}(t') = 0$ [see eq.~(\ref{v10})],
so that $v_{\text{net}}$ is given by
\begin{equation}
v_{\text{net}} = \alpha\tau^{-1} \int_0^\tau \mbox{d}t' \, v_{1,1}(t') \, .
\label{vnetlo}
\end{equation}
Again, $v_{\text{net}}$ is zero when $\alpha = 0$,
consistent with the notion that no directed motion occurs for symmetrical particles.

It is straightforward to recover the net velocity for
square forcing to lowest order in $\alpha$, eq.~(\ref{lo}), from (\ref{vnetlo}).
Similarly, for harmonic driving,
$
f(t) = f_0 \sin(2\pi t/\tau),
$
we find a time-average net velocity
\begin{equation}
v_{\text{net}} = \frac{\alpha f_0^2}{2} \frac{\tau^2}{4\pi^2 + \tau^2}\, .
\label{loh}
\end{equation}
For sawtooth forcing,
\begin{align}
0 \le t < \tau/2 :    \quad &f(t) = f_0 \,(t-\tau/4)/(\tau/4) \, , \nonumber \\
t \le \tau/2 < \tau : \quad &f(t) = f_0 \,(3\tau/4-t)/(\tau/4) \, ,
\end{align}
a net speed
\begin{equation}
v_{\text{net}} = \frac{\alpha f_0^2}{3} \left[ 1 - 3 \left(\frac{4}{\tau} \right)^2 + 3 \left(\frac{4}{\tau} \right)^3 \tanh \frac{\tau}{4} \right]
\label{los}
\end{equation}
is obtained.
These first order results for the net velocity $v_{\text{net}}$ are compared with the analytical
solution for square forcing in fig.~\ref{fig:netvelocity} (dashed
curves). We conclude that the resulting drift behavior is very similar
in all three cases. In fact, comparing eqs.~(\ref{lo}), (\ref{loh}) and (\ref{los}),
we see that
the differences become very small, and even vanish for slow forcing $\tau \to \infty$, if,
instead of using the same amplitude for the three modulations, one
considers the same average quadratic amplitude, \textit{i.e.}, if one replaces
$f_0/\sqrt{2} \to f_0$ in eq.~(\ref{loh}) and $f_0/\sqrt{3} \to f_0$ in  eq.~(\ref{los}).
\section{Microscopic models}
As already mentioned, the structure of eq. (\ref{equationmotion})
can be obtained from kinetic theory of microscopic models, that describe
a small, non-trivially shaped object (mass $M$)
moving under the influence of collisions with a
surrounding bath of gas particles (mass $m$), by an expansion in the
mass ratio $\varepsilon=\sqrt{m/M}$
\cite{vandenbroeckmotor, meurs, martijn1}.
Such a procedure also provides explicit expressions for the parameters
$\tau_r$ and $\alpha$ of the intrinsic ratchet. 

For a three-dimensional asymmetric object of arbitrary
convex shape, that is confined to move along a fixed $z$-axis
(translational motion), one obtains \cite{martijn1}
\begin{align}
\tau_r &= \varepsilon^{-2} \sigma_2^{-1}, \label{taumicr}\\
\alpha &=  \sqrt{\pi / 8}\, \varepsilon^3 \tau_r \sigma_3, \label{alphamicr}
\end{align}
where the geometry dependent moments $\sigma_n$ are given by
\begin{equation}
\sigma_n =  \rho \sqrt{\frac{8 k_B T}{\pi m}} \int_{\text{S}} \upd S \left(-\vec{e}_\perp |_z\right)^n,
\label{sigma}
\end{equation}
with $\rho$ being the particle density of the gas, and where the integral is
over the surface of the asymmetric object.  $\vec{e}_\perp |_z$ is the
component in the free direction of motion ($z$) of the outward unit
normal vector $\vec{e}_\perp$ at its surface.

\begin{figure}
\onefigure[width=0.75\columnwidth]{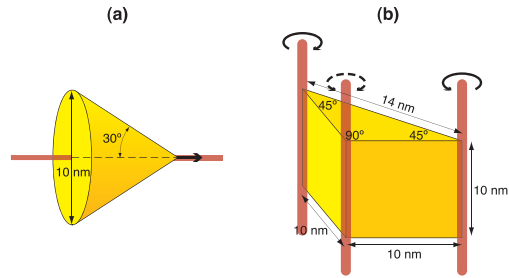}
\caption{Idealized realizations of the intrinsic ratchet. \textbf{(a)}
For translational motion: a conical shape with axis along the free
direction of motion. Indicated is the resulting sense of net motion
under unbiased forcing.  \textbf{(b)} For rotational motion: a right
triangular prism, with three suggested locations for the rotation
axis. The resulting net rotation sense under unbiased forcing is
indicated.}
\label{fig:realizations}
\end{figure}
To get an idea of the actual net velocity a realistic setup of an
intrinsic ratchet can attain, we use
eqs.~(\ref{lo}), (\ref{taumicr})-(\ref{sigma}) to calculate the speed
for a cone-shaped silica (\chem{SiO_2}) Brownian particle with half
opening angle 30$^\circ$ and 10\,nm base diameter [see
fig.~\ref{fig:realizations}(a) for a schematic representation]. The
cone's axis is along the free direction of motion. In air, the ratio
$M/m$ is about $10\,000$ and the asymmetry parameter is $\alpha =
0.003$. At temperature $T = 300\,$K the relaxation time is $\tau_r =
7.5\,$ns. An amplitude $f_0=10$ of unbiased square forcing then
corresponds to 1.9\,pN in real units and is well within the accuracy
range of our theory: $\alpha f_0 = 0.03$. These conditions produce a
maximum speed of $v_{\text{net}}^{\text{lim}} = 0.88\,$m/s or 30\% of
the thermal speed of the particle. Note that the direction of the
particle's motion is towards the apex of the cone.
\section{Rotational Brownian motion}
For simplicity of presentation, we started with the generic equations
of motion for one-dimensional translational Brownian motion of an
asymmetric particle. In practice, this
supposes that the particle is
constrained to move on a track.
The so-far presented discussion of the intrinsic ratchet can however be
repeated, with minor modifications, for the \emph{rotational} Brownian motion
of chiral objects, with angular velocity $\omega$ and moment of
inertia $I$.  With an adaptation of the expressions for the relaxation
time $\tau_r = I/\gamma$ and the thermal velocity $v_T = \sqrt{k_B
T/I}$ as units of time and angular velocity, and $f$ now signifying
a torque, this leads to the same generic equations of motion
[eq.~(\ref{equationmotion})] for the moments $v_1=\langle \omega
\rangle$ and $v_2=\langle \omega ^{2} \rangle-1$.  Microscopic theory
\cite{martijn2,martijn3} yields the same expressions for the
relaxation time $\tau_r$ [eq.~(\ref{taumicr})] and asymmetry
coefficient [eq.~(\ref{alphamicr})], but now with geometrical moments
given by
\begin{equation}
\sigma_n =  \rho \sqrt{\frac{8 k_B T}{\pi m}} \int_{\text{S}} \upd S
\left[(\vec{e}_\perp \times \vec{e}_r)|_z\right]^n\, ,
\label{momentrot}
\end{equation}
where the axis of rotation is taken to be parallel to the $z$-axis.
Again, the integral is over the surface of the object and
$\vec{e}_\perp$ is the outward unit normal vector on the surface.
$\vec{e}_r$ is given by $\vec{r} / r_0$, with $\vec{r}$ denoting the
position of a surface element measured from the axis of rotation
(the $z$-component in $\vec{r}$ is irrelevant),
and $r_0 = \sqrt{I / M}$ being the
radius of gyration of the object. Due to the chosen orientation of the
rotation axis, only the $z$-component of
$\vec{e}_\perp \times \vec{e}_r$
appears in the expression for $\sigma_n$.  With these new notations
and units, the previous results, in particular the expressions for the
time-average net velocity [eqs.~(\ref{lo}), (\ref{loh}), (\ref{los})], remain valid.

In view of the technological potential of the rotational setup, and
in order to get an idea of the order of magnitudes involved, we again
consider a realistic physical realization. A silica triangular prism
of height 10\,nm and with right triangular top and bottom surfaces
(sides: 10\,nm, 10\,nm, 14\,nm) is connected with a rotation axis at
one of its vertical edges, \textit{cf.}
fig.~\ref{fig:realizations}(b). Operating in air, the ratio $M/m$ is
about $23\,000$. If the axis is connected to the $90^\circ$ corner edge,
$\alpha = 0$ and no rectification or net rotation will
occur. Connected to the $45^\circ$ corner edges, $\alpha = \pm 0.0016$
and relaxation time is $\tau_r = 8.1\,$ns, at air temperature $T =
300\,$K.  A torque amplitude (for square forcing) of $f_0=10$
corresponds to $2.2\times 10^{-20}\,$Nm and produces a maximum net
rotation frequency of 6 MHz, 16\% of the thermal frequency.

\section{Molecular dynamics simulations}
In the following we verify our generic theory for the intrinsic
ratchet with molecular dynamics simulations.
As concrete microscopic system, we consider
the prism
from fig.~\ref{fig:realizations}(b) surrounded by a
thermalized bath of ideal gas particles, and perform event-driven
simulations of its rotational Brownian motion.
The rotation axis is chosen to be located slightly [$4\,\mbox{nm}$ in
the units of fig.~\ref{fig:realizations}(b)] outside the prism in the
plane given by one of the prism surfaces merging at the $90^\circ$
edge, and is oriented parallel to this edge.

Exploiting the homogeneity of the prism along the direction of the
rotation axis, the simulations are carried out in a (projected)
two-dimensional space, where the ratchet object is given by the right
triangular top (or bottom) surface of the prism, and where the rotation axis is
reduced to a point-like center of rotation. The rotation center
is positioned at the center of a quadratic box,
containing an ideal gas of point particles (mass $m$). The box walls `absorb' gas
particles upon collision, but also randomly `emit' new particles (into the
box' interior) such that
the gas properties, in
particular density $\rho$ and Maxwellian equilibrium distribution, are preserved.
In this way, an infinitely large reservoir of gas particles is
realized.

Collisions between gas particles and the triangle are detected by
numerically solving the exact equations of motion for the point in
time of the impact.
At each collision, the speed of the gas particle and the rotational
velocity of the triangle are changed according to the rules
for elastic collisions, neglecting tangential forces \cite{martijn2}. In
between collisions the triangle is accelerated by an external constant
or periodically switching torque (square profile).

\begin{figure}
\onefigure[width=0.9\columnwidth]{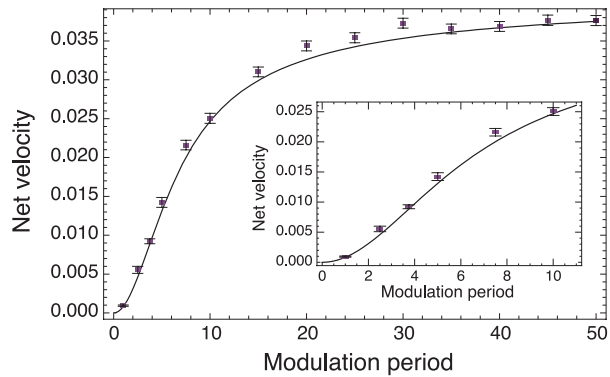}
\caption{Molecular dynamics simulation results (dots with error bars) compared with
theory (curves) for the rotational intrinsic ratchet. Time-average
net velocity $v_{\text{net}}$ is shown as a function of the
modulation period $\tau$ of zero-average square forcing. The simulation
results are obtained from averaging over typically $5000$
realizations ($20\,000$ for the smallest driving periods) with about
$17\,000$ collisions performed per realization, corresponding to a simulation
time of $1000$ periods for the fastest and $20$
periods for the slowest modulation; the error bars characterize the
remaining statistical uncertainty. Simulation parameters
are $\rho=0.25$, $M=50$, $m=1$, side lengths $(1,1,\sqrt{2})$ of the
right triangle, and amplitude $0.15$ of the external
square torque (see also main text), resulting in $\alpha = 0.0233$ and $f_0 =
1.32$. Units are relaxation time $\tau_r$ (time), thermal angular
velocity $v_T=\sqrt{k_B T/I}$ (angular velocity) and $I v_T /
\tau_r$ (torque), with $I$ the inertial moment.}
\label{fig:simulation}
\end{figure}

In fig.~\ref{fig:simulation} the net rotational speed of the triangle
under the action of a periodically switching torque is shown for
different values of the modulation period $\tau$. The agreement between simulation results for
the time-average net
velocity $v_{\text{net}}$ and theory, eq.~(\ref{lo}), is excellent.
We also compared the asymptotic rotation of the triangle  
under constant but opposite torques (`infinite' driving period
$\tau$) with the theoretical result, eq.~(\ref{steadystate}), and again
found excellent agreement.

\section{Stop-and-go motor}
\begin{figure}
\onefigure[width=0.9\columnwidth]{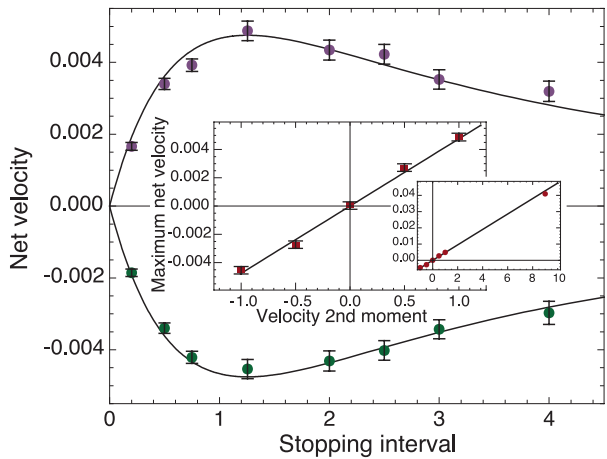}
\caption{Main figure: average (angular) net velocity $v_{\text{net}}$ of the
rotating stop-and-go ratchet as a function of the time interval $\tau_s$
between stopping events. In the trapping mechanism the second moment is
set to $v_{2,s} = 1$ (upper curve), $v_{2,s} = -1$ (lower curve),
with $v_2=\langle \omega^{2} \rangle-1$.  Insets: maximal average net
velocity $v_{\text{net}}$ (for optimal $\tau_s$) as a function of the
second moment of the velocity $v_{2,s}$ in the trap.  Lines
correspond to theory, dots represent results from molecular dynamics
simulations, using the same setup as for
fig.~\ref{fig:simulation} (but without external torque). The net speed is obtained from
simulating $1000$ independent stopping intervals $\tau_s$ per
realization and averaging over $10\,000$ realizations for the three smallest
stopping intervals and over $5000$ realizations otherwise.  
Shown angular velocity is in units of thermal speed $v_T=\sqrt{k_B T/I}$ and time
in units of relaxation time $\tau_r$.  The asymmetry parameter is
$\alpha = 0.0233$.}
\label{fig:stopandgo}
\end{figure}
We finally discuss an alternative approach to intrinsic ratchets,
anticipated in \cite{handrich} and further worked out in more detail
in \cite{sporer}. The so-called stop-and-go motor consists of an
asymmetric particle which is periodically stopped, for example by an
array of traps or binding sites that can be activated or deactivated
at will. The basic
assumption is that the (thermal) energy of the Brownian particle is changed
when it is subjected to the trapping mechanism. This energy exchange
results in a specific value 
of the second velocity moment at the stopping sites, $v_{2,s}$.
For $v_{2,s} \neq 0$
the energy exchange in the traps induces a deviation from
thermal equilibrium conditions, and this process will
result in sustained directed motion with an average net speed
$v_{\text{net}}$, being the average distance traveled by the particle
in a time interval $\tau_s$ between the stopping events,
divided by $\tau_s$. A simple analytical
calculation, starting from eq.~(\ref{equationmotion}) with $f = 0$,
gives an exact expression:
\begin{equation}
v_{\text{net}} = \frac{\alpha v_{2,s}}{2\tau_s} \left(1 - e^{-\tau_s}\right)^2.
\end{equation}
The sense of motion is determined by the sign of $\alpha$ and of
$v_{2,s}$, which is negative (positive) when the particle's thermal
motion is reduced (enhanced) by the trapping mechanism.
A stopping interval $\tau_s^o \simeq
1.26$, given by the solution of $e^{\tau_s} = 2\tau_s+1$, yields a
maximum net velocity ($v_{\text{net}}^{\text{max}} \simeq
0.204\,\alpha v_{2,s}$) and an optimal distance between binding sites
($\simeq 0.256\,\alpha v_{2,s}$, expressed in units $v_T \tau_r$).

These theoretical predictions are confirmed in a molecular dynamics simulation of the
stop-and-go mechanism applied to a rotating (chiral) object,
using the setup based on the prism in fig.~\ref{fig:realizations}(b),
as in the previous section.
In fig.~\ref{fig:stopandgo} the resulting average net velocity
$v_{\text{net}}$ as a function of different stop intervals $\tau_s$,
for the values $v_{2,s} = -1$ and 1 is shown. In the insets of
fig.~\ref{fig:stopandgo} we include the molecular dynamics results for
$v_{\text{net}}$ at the optimal stopping interval as a function of
$v_{2,s}$. The linear relation ($v_{\text{net}}^{\text{max}} \simeq
0.204\,\alpha v_{2,s}$) holds, even for large $v_{2,s}$.
\section{Conclusion}
Intrinsic ratchets are characterized by an inherent asymmetry
of the Brownian particle itself breaking
the spatial symmetry.
A generic formalism for the dynamical behavior
enables us to quantify the net particle velocity under unbiased periodic
forcing.
Molecular dynamics simulations of a rotational setup confirm the
validity of this formalism.
We predict drift speeds comparable to thermal
speeds for nanosized asymmetric Brownian particles under ratchet
operation.
The relative simplicity of the setup (one heat bath, external
symmetric forcing) could open avenues to experimentally test the
nonlinear contribution of intrinsic asymmetry crucial in this and
other work \cite{vandenbroeckmotor, meurs, martijn1, martijn2, martijn3,
 handrich, sporer, cleuren, jstat, costantini}.

\acknowledgments MvdB thanks Peter Reimann and his Condensed Matter
Theory Group of the University of Bielefeld for the kind hospitality
during numerous visits.
RE acknowledges support by the Deutsche
Forschungsgemeinschaft (SFB613).

\end{document}